\documentstyle[12pt,epsf]{article}

\newcommand{\ie}{{\it i.e.}}

\newcommand{\ms}{$\overline{\mbox{MS}}$}
\newcommand{\ams}{\mbox{$\alpha_{\overline{\mbox{\tiny MS}}}$}}
\newcommand{\amst}{\mbox{${\widetilde{\alpha}_{\overline{\mbox{\tiny MS}}}}$}}
\newcommand{\av}{\mbox{$\alpha_{V}$}}

\renewcommand{\bar}{\overline}

\newcommand{\GeV}{{\rm GeV}}

\def\npb#1#2#3{{\it Nucl. Phys. }{\bf B #1} (#2) #3}
\def\plb#1#2#3{{\it Phys. Lett. }{\bf B #1} (#2) #3}
\def\prd#1#2#3{{\it Phys. Rev. }{\bf D #1} (#2) #3}

\begin{document}
\begin{flushright}
SLAC--PUB--8022 \\
December 1998
\end{flushright}
\vfill
\begin{center}
{\large{Scheme-Independent Predictions in QCD:  Commensurate Scale
Relations and Physical Renormalization Schemes}
\footnote{\baselineskip=13pt Work partially supported by the Department
of Energy, contract DE--AC03--76SF00515.}}

\vspace{15mm}
{\bf S. J. Brodsky and J. Rathsman}\\
\vspace{5mm}
{\em Stanford Linear Accelerator Center \\
               Stanford University, Stanford, California 94309, USA \\
               e-mail:  sjbth@slac.stanford.edu -- rathsman@slac.stanford.edu}
\end{center}
\vfill
\begin{center}
Contribution to the Proceedings of the \\
IVth INTERNATIONAL SYMPOSIUM ON
RADIATIVE CORRECTIONS\\

Universitat Autonoma de Barcelona\\
Barcelona (Catalonia, Spain)\\
September 8--12, 1998\\
\vfill
\end{center}
\newpage

\hbox{ }

$$ $$

\begin{center}
Abstract
\end{center}

Commensurate scale relations are perturbative QCD predictions  which
relate observable to observable at fixed relative scale, such as the 
``generalized Crewther relation", which connects the Bjorken and 
Gross-Llewellyn Smith deep inelastic scattering sum rules to measurements 
of the $e^+ e^-$ annihilation cross section.  All non-conformal effects are 
absorbed by fixing the ratio of the respective momentum transfer and
energy scales. In the case of fixed-point theories, commensurate scale 
relations relate both the ratio of couplings and the ratio of scales as the fixed 
point is approached. The relations between the observables are independent 
of the choice of intermediate renormalization scheme or other theoretical 
conventions. Commensurate scale relations also provide an extension of the 
standard minimal subtraction scheme, which is analytic in the quark masses, 
has non-ambiguous scale-setting properties, and inherits the physical properties 
of the effective charge $\alpha_V(Q^2)$ defined from the heavy quark potential.  
The application of the analytic scheme to the calculation of quark-mass-dependent 
QCD corrections to the $Z$ width is also reviewed.

\vfill
\newpage

\section{Introduction}

One of the central problems in constructing precision tests of
a quantum field theory such as quantum chromodynamics is the elimination of
theoretical ambiguities such as the dependence on the renormalization
scale $\mu$ in perturbative expansions in the coupling $\alpha_s(\mu)$.
However, any prediction which
relates one physical quantity to another cannot depend on theoretical
conventions such as the choice of renormalization scheme or renormalization scale.   
This is the principle underlying ``commensurate scale relations" (CSR) \cite{CSR}, 
which are general QCD predictions relating physical observables to each other.
For example, the ``generalized Crewther relation",  which is discussed in more 
detail below,  provides a scheme-independent relation between the QCD 
corrections to the Bjorken (or Gross Llewellyn-Smith) sum rule for deep inelastic 
lepton-nucleon scattering,  at a given momentum transfer $Q$, to the radiative 
corrections to the annihilation cross section $\sigma_{e^+ e^- \to
\rm hadrons}(s)$, at a corresponding ``commensurate" energy scale $\sqrt s$.
\cite{CSR,BGKL}   The specific relation between the physical scales  $Q$
and $\sqrt s$ reflects the fact that the radiative corrections to each process have
distinct quark mass thresholds.

The generalized Crewther relation can be derived by calculating the
QCD radiative corrections to the deep inelastic sum rules and $R_{e^+ e^-}$ in a
convenient renormalization scheme such as the modified minimal subtraction
scheme $\overline{\rm MS}$. One then algebraically eliminates $\alpha_{\overline
{MS}}(\mu)$.
Finally, BLM scale-setting \cite{BLM} is used to eliminate the $\beta$-function
dependence of the coefficients.  The form of the resulting relation between the
observables thus matches the result which would have been obtained had QCD
been a
conformal theory with zero $\beta$ function.   The final result  relating the
observables is independent of the choice of intermediate
$\overline{\rm MS}$ renormalization scheme.

In quantum electrodynamics, the  running coupling $\alpha_{QED}(Q^2)$,
defined from
the Coulomb scattering of two heavy test charges at the momentum transfer
$t = -Q^2$, is
taken as the standard observable.  Similarly,  one can take the
momentum-dependent coupling
$\alpha_V(Q^2)$, defined from the potential scattering for heavy color
charges, as a
standard QCD observable.  Commensurate scale relations between $\alpha_V$
and the QCD
radiative corrections to other observables have no scale or scheme
ambiguity, even in
multiple-scale problems such as multijet production.   As is the case in
QED, the
momentum scale which appears as the argument of
$\alpha_V$ reflect the mean virtuality of the exchanged gluons.
Furthermore, we can
write a commensurate scale relation between $\alpha_V$ and an analytic
extension of
the
$\alpha_{\overline {MS}}$ coupling, thus transferring all of the
unambiguous scale-fixing
and analytic properties of the physical $\alpha_V$ scheme to the $\overline {MS}$
coupling.

Commensurate scale relations  thus provide
fundamental and precise scheme-independent tests of QCD, predicting  how
observables
track not only in relative normalization, but also in their commensurate scale
dependence.

\section{The Generalized Crewther Relation}

Any perturbatively calculable physical quantity can be used to define an
effective charge \cite{Grunberg,DharGupta,GuptaShirkovTarasov} by incorporating
the entire radiative correction into its definition. All such effective charges
$\alpha_A(Q)$ satisfy the Gell-Mann-Low renormalization group
equation.  In the case of massless quarks, the first two terms in the
perturbative
expansion for the
$\beta$ function of each effective charge,
$\beta_0$ and $\beta_1$, are universal;  different schemes or effective
charges only
differ through the third and higher coefficients. Any
effective charge can be used as a reference running coupling constant in
QCD to define
the renormalization
procedure.  More generally, each effective charge or renormalization
scheme, including
$\overline{\rm MS}$, is a special case of the universal coupling function
$\alpha(Q, \beta_n)$.

For example, consider the Adler function \cite{Adler} for the $e^+ e^-$
annihilation cross section
\begin{equation} D(Q^2)=-12\pi^2 Q^2{d\over dQ^2}\Pi(Q^2),~
\Pi(Q^2) =-{Q^2\over 12\pi^2}\int_{4m_{\pi}^2}^{\infty}{R_{e^+ e^-}(s)ds\over
s(s+Q^2)}.
\end{equation}
The entire radiative correction to this function is defined as
the effective charge
$\alpha_D(Q^2)$ :
\begin{eqnarray}
    D \left( Q^2/ \mu^2, \alpha_{\rm s}(\mu^2) \right) &=&
D \left (1, \alpha_{\rm s}(Q^2)\right) \label{3} \\
&\equiv&
    3 \sum_f Q_f^2 \left[ 1+ {3\over 4} C_F{\alpha_D(Q^2) \over \pi}
                   \right]
    +( \sum_f Q_f)^2C_{\rm L}(Q^2) \nonumber \\
&\equiv& 3 \sum_f Q_f^2 C_D(Q^2)+( \sum_f Q_f)^2C_{\rm L}(Q^2),
\nonumber
\end{eqnarray}
where $C_F={N_C^2-1\over 2 N_C}. $
The coefficient $C_{\rm L}(Q^2)$ appears at the third order in
perturbation theory and is related to the ``light-by-light scattering type"
diagrams.  (Hereafter $\alpha_{\rm s}$ will denote the ${\overline{\rm MS}}$
scheme strong coupling constant.)
Similarly, we can define the entire radiative correction to the Bjorken sum rule
as the effective charge $\alpha_{g_1}(Q^2)$ where
$Q$ is the corresponding momentum transfer:
\begin{equation}
\int_0^1 d x \left[ g_1^{ep}(x,Q^2) - g_1^{en}(x,Q^2) \right]
   \equiv {1\over 6} \left|g_A \over g_V \right|
   C_{\rm Bj}(Q^2)
 = {1\over 6} \left|g_A
\over g_V \right| \left[ 1- {3\over 4} C_F{\alpha_{g_1}(Q^2) \over
   \pi} \right]  .
\end{equation}
It is straightforward to algebraically relate $\alpha_{g_1}(Q^2)$ to
$\alpha_D(Q^2)$ using the known expressions to three loops in the
$\overline{\rm MS}$ scheme. If one chooses the renormalization scale to resum
all of the quark and gluon vacuum polarization corrections into
$\alpha_D(Q^2)$,  then
the final result turns out to be remarkably simple \cite{BGKL}
 $(\widehat\alpha = 3/4\, C_F\ \alpha/\pi):$
\begin{equation}
\widehat{\alpha}_{g_1}(Q)=\widehat{\alpha}_D( Q^*)-
\widehat{\alpha}_D^2( Q^*)+\widehat{\alpha}_D^3( Q^*) +
\cdots,
\end{equation}
where
\begin{eqnarray}
\ln \left({ {Q}^{*2} \over Q^2} \right) &=&
{7\over 2}-4\zeta(3)+\left(\frac{\alpha_D ( Q^*)}{4\pi}
\right)\Biggl[ \left(
            {11\over 12}+{56\over 3} \zeta(3)-16{\zeta^2(3)}
      \right) \beta_0\cr &&
      +{26\over 9}C_{\rm A}
      -{8\over 3}C_{\rm A}\zeta(3)
      -{145\over 18} C_{\rm F}
      -{184\over 3}C_{\rm F}\zeta(3)
      +80C_{\rm F}\zeta(5)
\Biggr].
\label{EqLogScaleRatio}
\end{eqnarray}
where in QCD, $C_{\rm A}=N_C = 3$ and $C_{\rm F}=4/3$.
This relation shows how
the coefficient functions for these two different processes are
related to each other at their respective commensurate scales. We emphasize
that the $\overline{\rm MS}$ renormalization
scheme is used only for calculational convenience; it serves simply as an
intermediary between observables. The renormalization group
ensures
that the forms of the CSR relations in perturbative QCD are independent
of the choice of an intermediate renormalization scheme.

The Crewther relation was originally derived assuming that the theory is
conformally
invariant; \ie, for zero $\beta$  function. In the physical case, where the
QCD coupling runs,  all non-conformal effects are resummed into the
energy and momentum transfer scales of
the effective couplings $\alpha_R$ and $\alpha_{g1}$.  The general
relation between these two effective charges for nonconformal
theory  thus takes the form of a geometric series
\begin{equation}
        1- \widehat \alpha_{g_1} =
\left[ 1+ \widehat \alpha_D( Q^*)\right]^{-1} \ .
\end{equation}
We have dropped the small light-by-light scattering contributions.
This is again a special advantage of relating observable to
observable.
The coefficients are independent of
color and are the same in Abelian, non-Abelian, and conformal gauge theory.  The
non-Abelian structure of the theory is reflected in the expression for the scale
${Q}^{*}$.

Is experiment consistent with the generalized Crewther relation? Fits
\cite{MattinglyStevenson} to the experimental measurements of the
$R$-ratio above the thresholds for the production of $c\overline{c}$ bound
states
provide the empirical constraint:
$\alpha_{R}({\sqrt s} =5.0~{\rm GeV})/\pi  \simeq 0.08\pm 0.03.$
The prediction for the effective coupling for
the deep inelastic sum rules at the commensurate momentum transfer $Q$
is then
$\alpha_{g_1}(Q=12.33\pm 1.20~{\rm GeV})/\pi
\simeq \alpha_{\rm GLS}(Q=12.33\pm 1.20~{\rm GeV})/\pi
\simeq 0.074\pm 0.026.$
Measurements of the Gross-Llewellyn Smith sum rule have so far only been
carried out
at relatively small values of $Q^2$ \cite{CCFRL1,CCFRL2};
however, one can use the results of the theoretical
extrapolation \cite{KS} of the experimental data presented in \cite{CCFRQ}:
$ \alpha_{\rm GLS}^{\rm extrapol}(Q=12.25~{\rm GeV})/\pi\simeq 0.093\pm 0.042.$
This range  overlaps with the prediction from  the generalized
Crewther relation.   It is clearly important to have higher precision
measurements to
fully test this fundamental QCD prediction.

\section{General Form of Commensurate Scale Relations}

In general, commensurate scale
relations connecting the effective charges for observables $A$ and $B$ have the
form
\begin{equation}
\alpha_A(Q_A) =
\alpha_B(Q_B) \left(1 + r^{(1)}_{A/B} {\alpha_B(Q_B)\over \pi} +  r^{(2)}_{A/B}
{\alpha_B(Q_B)\over
\pi}^2 + \cdots\right),
\label{eq:CSRg}
\end{equation}
where the coefficients $r^{{n}}_{A/B}$
are
identical to the coefficients obtained in a con\-formally invariant theory
with $\beta_B(\alpha_B) \equiv (d/d\ln Q^2) \alpha_B(Q^2) = 0$.
The ratio of the scales $Q_A/Q_B$ is thus fixed by the requirement that
the couplings sum all of the effects of the non-zero $\beta$ function.   In
practice the
NLO and NNLO coefficients and relative scales can be identified from the flavor
dependence of the perturbative series; \ie\  by shifting scales such that the
$N_F$-dependence associated with $\beta_0 = 11/3 C_A - 4/3 T_F N_F$ and
$\beta_1 =
-34/3 C_A^2 + {20\over 3} C_A T_F N_F + 4 C_F T_F N_F$
does not appear in
the coefficients. Here $C_A=N_C$, $C_F=(N^2_C-1)/2N_C$ and $T_F=1/2$.
The shift in scales which gives conformal coefficients in effect pre-sums
the large and strongly divergent terms in the PQCD  series which grow as
$n! (\beta_0
\alpha_s)^n$, \ie, the infrared renormalons associated with coupling-constant
renormalization. \cite{tHooft,Mueller,LuOneDim,BenekeBraun}

The
renormalization scales $Q^*$ in the BLM method are physical in the sense that
they reflect the mean virtuality of the gluon propagators.  This
scale-fixing procedure is consistent with scale fixing in  QED, in agreement
with
in the Abelian limit, $N_C \to 0$. \cite{BrodskyHuet}
\cite{BLM,LepageMackenzie,Neubert,BallBenekeBraun}
The ratio of scales
$\lambda_{A/B} = Q_A/Q_B$  guarantees that the
observables $A$ and $B$ pass through new quark thresholds at the same physical
scale.  One can also show that the commensurate scales satisfy the
transitivity rule
$\lambda_{A/B} = \lambda_{A/C} \lambda_{C/B},$  which ensures that predictions
are independent of the choice of an intermediate renormalization scheme or
intermediate observable $C.$

\section{Commensurate Scale Relations and Fixed Points}

In general, we can write the relation between any two effective charges at
arbitrary
scales
$\mu_A$ and
$\mu_B$ as a correction to the corresponding relation obtained in a
conformally invariant
theory:
\begin{equation}
\alpha_A(\mu_A) = C_{AB}[\alpha_B(\mu_B)] +
\beta_B[\alpha_B(\mu_B)] F_{AB}[\alpha_B(\mu_B)]
\label{eq:ak}
\end{equation}
where
\begin{equation}
 C_{AB}[\alpha_B] = \alpha_B + \sum_{n=1} C_{AB}^{(n)}\alpha^n_B
\label{eq:al}
\end{equation}
is the functional relation when $\beta_B[\alpha_B]=0$.  In fact, if $\alpha_B$
approaches a fixed  point $\bar\alpha_B$  where
$\beta_B[\bar\alpha_B]=0$,
then $\alpha_A$ tends to a fixed point given by
\begin{equation}
\alpha_A \to \bar\alpha_A =  C_{AB}[\bar\alpha_B].
\label{eq:am}
\end{equation}
The commensurate scale relation for observables $A$ and $B$ has a similar
form, but
in this case the relative scales are fixed such that the non-conformal term
$F_{AB}$ is
zero.
Thus the commensurate scale relation $\alpha_A(Q_A) = C_{AB}[\alpha_B(Q_B)]$ 
at general commensurate scales is also the relation connecting the values
of the fixed
points for any two effective charges or schemes.   Furthermore, as
$\beta\rightarrow 0$,
the ratio of commensurate scales $Q^2_A/Q^2_B$ becomes the ratio of
fixed point scales $\bar Q^2_A/\bar Q^2_B$ as one approaches
the fixed point regime.

\section{Implementation of  $\alpha_V$ Scheme}
\unboldmath

Is there a preferred effective charge which we should use to characterize the
coupling strength in QCD? In QED, the running coupling $\alpha_{QED}(Q^2)$,
defined from
the potential between two infinitely heavy test charges, has traditionally
played that
role.  In the case of QCD,  the heavy-quark potential $V(Q^2)$ is defined as the
two-particle-irreducible scattering amplitude of test color charges; \ie \ the
scattering of an infinitely heavy quark and antiquark at momentum transfer $t =
-Q^2.$  The relation $V(Q^2) = - 4 \pi C_F
\alpha_V(Q^2)/Q^2$ then defines
the effective charge $\alpha_V(Q).$  This coupling can provide a
physically based alternative to the usual ${\overline {MS}}$ scheme.
As in the corresponding case of Abelian QED, the scale $Q$ of the coupling
$\alpha_V(Q)$ is identified with the exchanged  momentum. Thus there is never
any ambiguity in the interpretation of the scale.   All vacuum polarization
corrections
due to fermion pairs are incorporated in  $\alpha_V$ through the usual
vacuum polarization
kernels which depend on the physical mass thresholds.  Of course, other
observables could be used to define the standard QCD coupling, such as the
effective
charge defined from heavy quark radiation. \cite{Uraltsev}

The relation of  $\alpha_V(Q^2)$  to the  conventional $\overline {MS}$
coupling  is now known to NNLO, \cite{Peter}
but in the following only the NLO relation will be used.  The commensurate
scale relation is given by  \cite{BGMR}
\begin{eqnarray}
\label{eq:csrmsovf}
\alpha_{\overline{\mbox{\tiny MS}}}(Q) 
& = & \alpha_V(Q^{*}) + \frac{2}{3}N_C{\alpha_V^2(Q^{*}) \over \pi}
\nonumber \\
& = & \alpha_V(Q^{*}) + 2{\alpha_V^2(Q^{*}) \over \pi}\  ,
\end{eqnarray} 
which is valid for $Q^2 \gg m^2$.  The coefficients in the
perturbation expansion have their conformal values, \ie, the same coefficients
would occur even if the theory had been conformally invariant with $\beta=0$.
The commensurate scale is given by
\begin{eqnarray} 
Q^* & = & Q\exp\left[\frac{5}{6}\right] \ .
\end{eqnarray}
 The scale in the $\overline {MS}$ scheme is thus a factor
$\sim 0.4$ smaller than the physical scale. The coefficient $2 N_C/3$ in the NLO
coefficient is a feature of the non-Abelian couplings of QCD; the same
coefficient occurs even if the theory were conformally invariant with
$\beta_0=0.$

Using the above QCD results, we can transform any NLO  prediction
given in $\overline{MS}$ scheme to a scale-fixed expansion in
$\alpha_V(Q)$.
We can also derive the connection between the $\overline{MS}$ and $\alpha_V$
schemes for  Abelian perturbation theory using  the limit $N_C \to 0$ with
$C_F\alpha_s$ and $N_F/C_F$ held fixed. \cite{BrodskyHuet}

The use of $\alpha_V$ and related physically defined effective charges such as
$\alpha_p$  (to NLO the effective charge defined from the (1,1) plaquette,
$\alpha_p$ is the same as $\alpha_V$) as expansion parameters  has been 
found to be  valuable in lattice
gauge theory, greatly increasing the convergence of perturbative expansions
relative to
those using the bare lattice coupling. \cite{LepageMackenzie}  Recent lattice
calculations of the
$\Upsilon$- spectrum \cite{Davies} have been used with BLM
scale-fixing to determine a NLO normalization
of the static heavy quark potential:  $
\alpha_V^{(3)}(8.2 \GeV) = 0.196(3)$ where the effective number of light
flavors is
$n_f = 3$.  The
corresponding modified minimal subtraction coupling evolved to the
$Z$ mass and five flavors is  $ \alpha_{\overline{MS}}^{(5)}(M_Z) =
0.1174(24)$. Thus a high precision value for $\alpha_V(Q^2)$  at a specific
scale is
available from lattice gauge theory.  Predictions for other QCD observables
can be
directly referenced to this value  without the scale or scheme ambiguities,
thus greatly
increasing the precision of QCD tests.

One can also use $\alpha_V$ to  characterize the coupling which appears in the hard
scattering contributions of exclusive process amplitudes at large momentum
transfer,
such as elastic hadronic  form factors, the photon-to-pion transition form
factor at
large momentum transfer \cite{BLM,BJPR}  and exclusive weak decays of heavy
hadrons.\cite{Henley}   Each gluon
propagator with four-momentum $k^\mu$ in the hard-scattering quark-gluon
scattering amplitude $T_H$ can be associated with the coupling
$\alpha_V(k^2)$ since the
gluon  exchange propagators closely resembles the interactions encoded in the
effective potential $V(Q^2)$. [In Abelian theory this is exact.]
Commensurate scale
relations can then be
established which connect the hard-scattering subprocess amplitudes which
control exclusive processes to other QCD observables.

We can anticipate that  eventually
nonperturbative
methods such as lattice gauge theory or discretized light-cone quantization will
provide a complete form for the heavy quark potential in $QCD$.   It
is reasonable to assume that $\alpha_V(Q)$ will not diverge at small space-like
momenta. One possibility is that $\alpha_V$ stays relatively constant
$\alpha_V(Q) \simeq 0.4$ at low momenta, consistent with  fixed-point behavior.
There is, in fact, empirical evidence for freezing of the $\alpha_V$  coupling
from the observed systematic dimensional scaling behavior of exclusive
reactions. \cite{BJPR}   If this is in fact the case, then the range of QCD
predictions can be extended to quite low momentum scales, a regime normally
avoided because of the apparent singular structure of perturbative
extrapolations.

There are a number of other advantages of the $V$-scheme:
\begin{enumerate}
\item
Perturbative expansions in $\alpha_V$ with the scale set by the momentum
transfer cannot
have any
$\beta$-function dependence in their coefficients  since all running
coupling effects are
already summed into the definition of the potential.   Since
coefficients involving
$\beta_0$ cannot occur in an expansions in $\alpha_V$,  the divergent infrared
renormalon series of the form $\alpha^n_V\beta_0^n n!$  cannot occur. The
general convergence properties of the scale $Q^*$ as an expansion in $\alpha_V$
is not known. \cite{Mueller}

\item
The effective coupling $\alpha_V(Q^2)$ incorporates vacuum polarization
contributions with finite fermion masses.  When continued to time-like
momenta, the coupling has the correct analytic dependence dictated by the
production thresholds in the $t$ channel. Since  $\alpha_V$ incorporates  quark
mass effects exactly, it avoids the problem of explicitly computing and
resumming
quark mass corrections.

\item
The  $\alpha_V$  coupling is the natural expansion parameter for
processes involving non-relativistic momenta, such as heavy quark production at
threshold where the Coulomb interactions, which are enhanced at low relative
velocity $v$ as $\pi \alpha_V/v$, need to be
re-summed. \cite{Voloshin,Hoang,Fadin}
The effective Hamiltonian for nonrelativistic QCD  is thus  most naturally
written in
$\alpha_V$ scheme.
The threshold corrections
to heavy quark production in $e^+ e^-$ annihilation depend on
$\alpha_V$ at specific scales $Q^*$.  Two distinct ranges of scales arise
as arguments of
$\alpha_V$ near threshold: the relative momentum of the quarks governing the
soft gluon exchange responsible for the Coulomb potential, and a high momentum
scale, induced by
hard gluon exchange, approximately equal to twice the quark mass for the
corrections.
\cite{Hoang}  One thus can use threshold production to obtain a direct
determination of $\alpha_V$ even at low scales.  The corresponding QED results
for $\tau$ pair production allow for a measurement of the magnetic moment of the
$\tau$ and could be tested at a future $\tau$-charm
factory. \cite{Voloshin,Hoang}

\end{enumerate}

We also note that
computations in different sectors of the Standard Model have been
traditionally carried out using different renormalization schemes.
However, in a grand
unified theory, the forces between all of the particles in the fundamental
representation should become universal above the grand unification scale.
Thus it is
natural to use $\alpha_V$ as the effective charge for all sectors of a grand
unified theory,  rather than in
a convention-dependent coupling such as $\alpha_{\overline {MS}}$.

\section{The Analytic Extension of the $\bar{MS}$ Scheme}

The standard ${\overline {MS}}$ scheme
is not an analytic function of the renormalization scale at heavy quark thresholds; 
in the running of the coupling the quarks are taken as massless, and
at each quark threshold the value of $N_F$ which appears in the $\beta$
function is
incremented. Thus Eq. (\ref{eq:csrmsovf}) is technically only valid far
above a heavy quark threshold.  However, we can use this commensurate scale
relation to define
an  extended
$\overline {MS}$  scheme which is continuous and analytic at any scale.  The new
modified scheme inherits all of the good properties of the $\alpha_V$ scheme,
including its correct analytic properties as a function of the quark masses
and its
unambiguous scale fixing. \cite{BGMR}
Thus we define
\begin{equation}
\widetilde {\alpha}_{\overline{\mbox{\tiny MS}}}(Q)
=  \alpha_V(Q^*) + \frac{2N_C}{3} {\alpha_V^2(Q^{**})\over\pi} +
\cdots ,
\label{alpmsbar2}
\end {equation}
for all scales $Q$.  This equation not only provides an analytic
extension of the $\overline{MS}$ and similar schemes, but it also ties down the
renormalization scale to the physical masses of the quarks as they
enter into the vacuum polarization contributions to $\alpha_V$.

The modified scheme \amst\ provides an analytic interpolation of
conventional $\overline{MS}$ expressions by utilizing the mass dependence of the
physical \av\ scheme. In effect, quark thresholds are treated
analytically to all orders in $m^2/Q^2$; \ie, the evolution of the analytically
extended coupling in the intermediate regions reflects the actual mass
dependence of a physical effective charge and the analytic properties of
particle production.
Just as in Abelian QED, the mass dependence of the effective potential
and the analytically extended scheme \amst\ reflects the analyticity of the
physical thresholds for particle production in the
crossed channel.  Furthermore, the definiteness of the dependence in the quark
masses automatically constrains the renormalization scale.  There
is thus no scale ambiguity in perturbative expansions in \av\ or \amst.

In leading order the effective number of flavors in the modified scheme
\amst\ is given to a very good approximation by the simple form \cite{BGMR}
\begin{equation}
\widetilde {N}_{F,\overline{\mbox{\tiny MS}}}^{(0)}\left(\frac{m^2}{Q^2}\right)
\cong \left(1 + {5m^2  \over {Q^2\exp({5\over 3})}} \right)^{-1}
\cong \left( 1 + {m^2  \over Q^2} \right)^{-1}.
\end{equation}
Thus the contribution from one flavor is $\simeq 0.5$ when
the scale $Q$ equals the quark mass $m_i$.  The standard procedure
of matching $\alpha_{\overline{\mbox{\tiny MS}}}(\mu)$ at the quark
masses serves as a zeroth-order approximation to the continuous $N_F$.

\begin{figure}[htb]
\begin{center}
\leavevmode
\epsfxsize=4in
\epsfbox{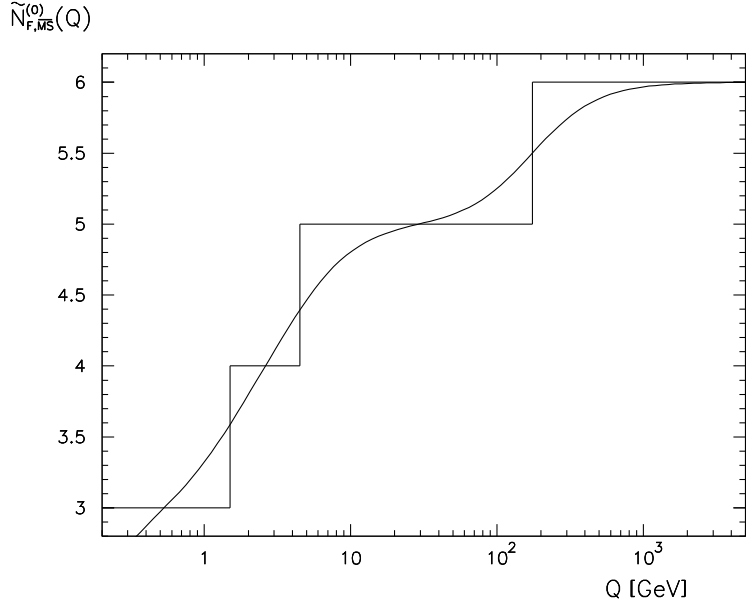}
\end{center}
\caption[*]{The continuous
$\widetilde {N}_{F,\overline{\mbox{\tiny MS}}}^{(0)}$ in the analytic
extension of the $\overline{\mbox{MS}}$ scheme as a
function of the physical scale $Q$. (For reference the
continuous $N_F$ is also compared with
the conventional procedure of taking $N_F$ to be a step-function at the
quark-mass thresholds.)}
\label{fig:nfsum}
\end{figure}

Adding all flavors together gives the total
$\widetilde {N}_{F,\overline{\mbox{\tiny MS}}}^{(0)}(Q)$
which is shown in Fig.~\ref{fig:nfsum}. For reference, the
continuous $N_F$ is also compared with
the conventional procedure of taking $N_F$ to be a step-function at the
quark-mass thresholds.
The figure shows clearly that there are hardly any plateaus at all
for the continuous
$\widetilde {N}_{F,\overline{\mbox{\tiny MS}}}^{(0)}(Q)$ in
between the quark masses.
Thus there is really no scale below 1 TeV where
$\widetilde {N}_{F,\overline{\mbox{\tiny MS}}}^{(0)}(Q)$
can be approximated by a constant; for all $Q$ below 1 TeV there is always
one quark
with mass $m_i$ such that $m_i^2 \ll Q^2$ or $Q^2 \gg m_i^2$ is not
true.
We also note that if one would use any other scale than the
BLM-scale for $\widetilde {N}_{F,\overline{\mbox{\tiny MS}}}^{(0)}(Q)$,
the result would be to increase the difference between the analytic
$N_F$ and the standard procedure of using the step-function at the
quark-mass thresholds.

\begin{figure}[htb]
\begin{center}
\leavevmode
\epsfxsize=4in
\epsfbox{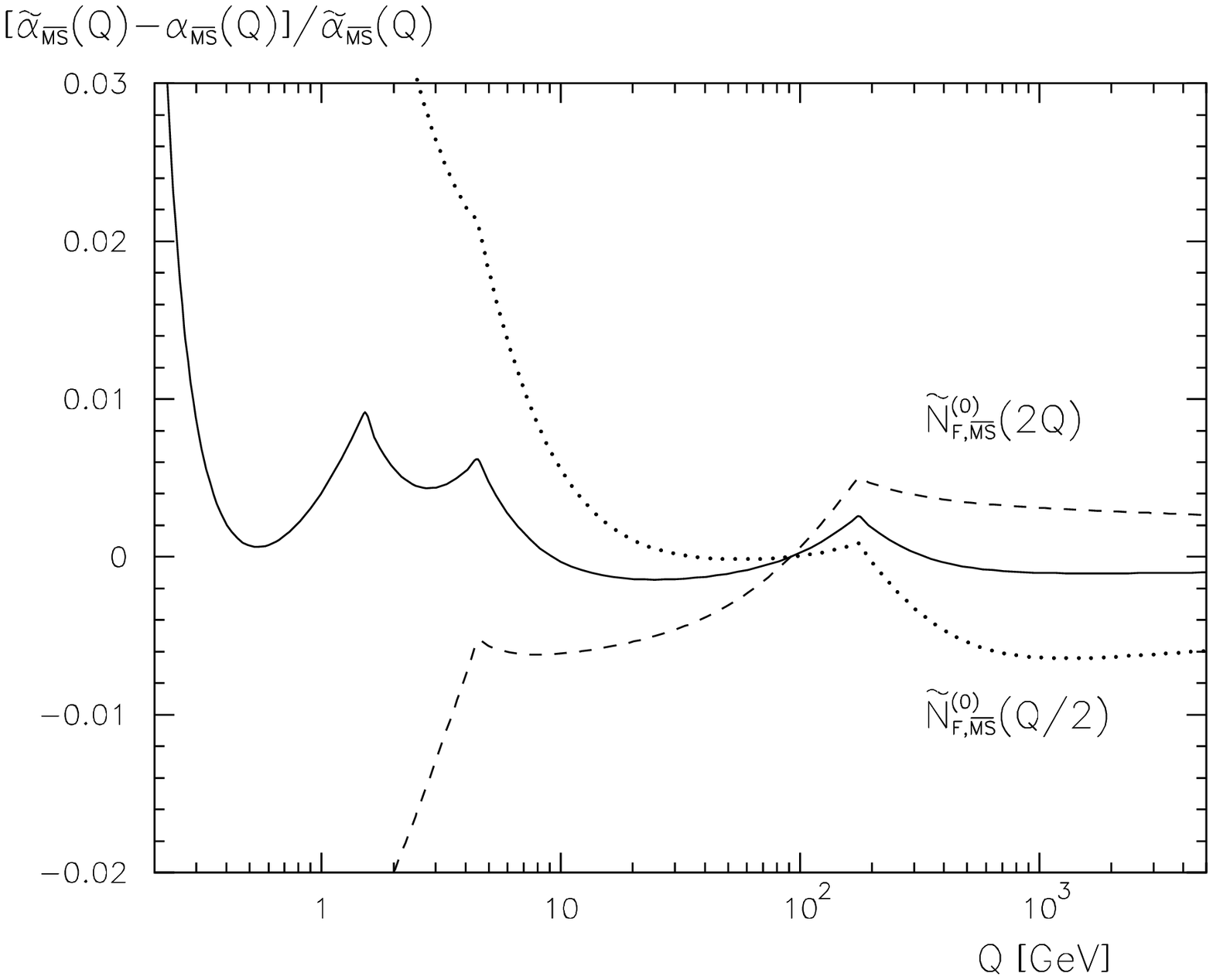}
\end{center}
\caption[*]{The solid curve shows the relative difference between the
solutions to
the 1-loop renormalization group equation using continuous $N_F$,
$\widetilde{\alpha}_{\overline{\mbox{\tiny MS}}}(Q)$, and conventional discrete
theta-function thresholds, $\alpha_{\overline{\mbox{\tiny MS}}}(Q)$.
The dashed (dotted) curves shows the same quantity but using the scale $2Q$
($Q/2$)
in $\widetilde {N}_{F,\overline{\mbox{\tiny MS}}}^{(0)}$.  The solutions
have been
obtained numerically starting from the world average \cite{Burrows}\
$\alpha_{\overline{\mbox{\tiny MS}}}(M_Z) = 0.118$.}
\label{fig:adiff}
\end{figure}

Figure~\ref{fig:adiff} shows the relative difference between the two
different solutions of the 1-loop renormalization group equation,
\ie\ $(\widetilde{\alpha}_{\overline{\mbox{\tiny MS}}}(Q)-
           {\alpha}_{\overline{\mbox{\tiny MS}}}(Q) )/
           \widetilde{\alpha}_{\overline{\mbox{\tiny MS}}}(Q)$.
The solutions have been obtained numerically starting from the
world average \cite{Burrows}
$\alpha_{\overline{\mbox{\tiny MS}}}(M_Z) = 0.118$.
The figure shows that
taking the quark masses into account in the running leads to
effects  of the order of one percent, most especially
pronounced near thresholds.

To illustrate how to compute an
observable using the analytic extension of the \ms\ scheme and compare
with the standard treatment in
the \ms\ scheme we consider the
QCD corrections to the quark part of the non-singlet hadronic width of
the Z-boson, $\Gamma_{had,q}^{NS}$. Writing the QCD corrections in terms
of an effective charge we have
\begin{equation}
\Gamma_{had,q}^{NS}=\frac{G_FM_Z^3}{2\pi\sqrt{2}}
\sum_{q}\{(g_V^{q})^2+(g_A^{q})^2\}
\left[1+\frac{3}{4}C_F\frac{\alpha_{\Gamma,q}^{NS}(s)}{\pi}\right]
\end{equation}
where the effective charge $\alpha_{\Gamma,q}^{NS}(s)$ contains all
QCD corrections,
\begin{eqnarray}
\frac{\alpha_{\Gamma,q}^{NS}(s)}{\pi} & = &
\frac{\alpha_{\overline{\mbox{\tiny MS}}}^{(N_L)}(\mu)}{\pi}
\Bigg\{1+\frac{\alpha_{\overline{\mbox{\tiny MS}}}^{(N_L)}(\mu)}{\pi}
\nonumber \\ && \times
\left[\sum_{q=1}^{N_L}\left(-\frac{11}{12}+\frac{2}{3}\zeta_3
+ F\left(\frac{m_q^2}{s}\right)
-\frac{1}{3}\ln\left(\frac{\mu}{\sqrt{s}}\right)\right)
 \right. \nonumber \\ && \left. \left.
+\sum_{Q=N_L+1}^{6}G\left(\frac{m_Q^2}{s}\right)\right] + \ldots \right\}
\end{eqnarray}

To calculate $\alpha_{\Gamma,q}^{NS}(s)$ in the
analytic extension of the \ms\ scheme one first applies
the BLM scale-setting procedure
in order to absorb all the massless effects of non-zero $N_F$ into the
running of the
coupling. This gives
\begin{eqnarray}
\label{eq:agms}
\frac{\alpha_{\Gamma,q}^{NS}(s)}{\pi} & = &
\frac{\alpha_{\overline{\mbox{\tiny MS}}}^{(N_L)}(Q^*)}{\pi}
\\ & & \times
\left\{1+\frac{\alpha_{\overline{\mbox{\tiny MS}}}^{(N_L)}(Q^*)}{\pi}
\left[\sum_{q=1}^{N_L}F\left(\frac{m_q^2}{s}\right)
+\sum_{Q=N_L+1}^{6}G\left(\frac{m_Q^2}{s}\right)\right] + \ldots \right\}
\nonumber
\end{eqnarray}
where
\begin{equation}
Q^*=\exp\left[3\left(-\frac{11}{12}+\frac{2}{3}\zeta_3\right)\right]\sqrt{s}
=0.7076\sqrt{s}.
\end{equation}
Operationally, one then  simply drops
all the mass dependent terms in the above expression and replaces the
fixed $N_F$ coupling $\alpha_{\overline{\mbox{\tiny MS}}}^{(N_L)}$
with the analytic \amst. (For an observable calculated with massless quarks
this step reduces to replacing the coupling.)
In this way both the massless $N_F$ contribution,
as well as the mass-dependent contributions from double bubble diagrams,
are absorbed into the coupling.
We are thus left with a very simple expression,
\begin{eqnarray}
\label{eq:aganalytic}
\frac{\alpha_{\Gamma,q}^{NS}(s)}{\pi} & = &
\frac{\amst(Q^*)}{\pi},
\end{eqnarray}
reflecting the fact that the QCD effects of quarks in the
perturbative coefficients, both massless and massive, should be absorbed
into the running of the coupling.

In order to compare the analytic extension  of the \ms\ scheme
with the standard \ms\ result for
$\alpha_{\Gamma,q}^{NS}(s)$, we will apply
the BLM scale-setting procedure also for the standard \ms\ scheme.
This is to ensure that any differences are due to the different
ways of treating quark masses and not due to the scale choice.
In other words we want to compare Eqs.~(\ref{eq:agms})
and (\ref{eq:aganalytic}). As the normalization point we use
$\alpha_{\overline{\mbox{\tiny MS}}}^{(5)}(M_Z)=0.118$
which we evolve down to $Q^*=0.7076M_Z$ using leading order
massless evolution with $N_F=5$. This value is then used to calculate
$\alpha_{\Gamma,q}^{NS}(M_Z)=0.1243$ in the \ms\ scheme using
Eq.~(\ref{eq:agms}). Finally, Eq.~(\ref{eq:aganalytic}) gives the
normalization point for $\amst(Q^*)$.

\begin{figure}[htb]
\begin{center}
\leavevmode
\epsfxsize=4in
\epsfbox{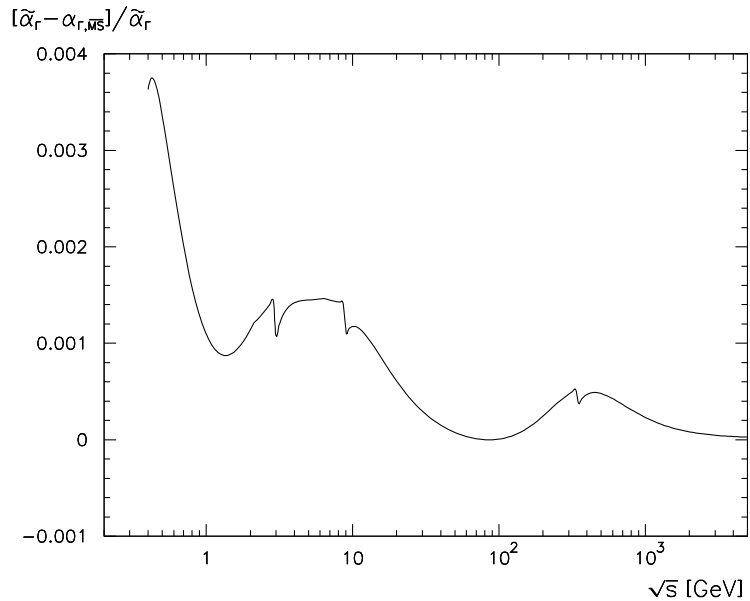}
\end{center}
\caption[*]{The relative difference between the calculation of
$\alpha_{\Gamma,q}^{NS}(s)$ in the analytic extension of the \ms\ scheme
and the standard treatment of masses in the \ms\ scheme. The discontinuities
are due to the mismatch between the $s/m^2$ and $m^2/s$ expansions of the
functions $F$ and $G$.}
\label{fig:gdiff}
\end{figure}

Figure~\ref{fig:gdiff} shows the relative difference between the two
expressions for $\alpha_{\Gamma,q}^{NS}(s)$ given by Eqs.~(\ref{eq:agms})
and (\ref{eq:aganalytic}) respectively. As can be seen from the figure the
relative difference is remarkably small, less than $0.2\%$ for scales above
1 GeV. Thus
the analytic extension of the \ms\ scheme takes the mass corrections
into account in a very simple way without having to include an infinite
series of higher dimension operators or doing complicated multi-loop
diagrams with explicit masses.

The form of $N_F(Q)$ at NNLO has recently been computed to two loop order
in QCD for the
$\alpha_V$ scheme. The application to the analytic extension of \ms\ scheme
will be
discussed in a forthcoming paper. \cite{Melles}

\section{Conclusion}

Commensurate scale relations have a number of attractive properties:
\begin{enumerate}
\item
The ratio of physical scales $Q_A/Q_B$ which appears in commensurate scale
relations
reflects the relative position of physical thresholds, \ie\, quark
anti-quark pair
production.
\item
The functional dependence and perturbative expansion of the CSR are identical to
those of a conformal scale-invariant theory where $\beta_A(\alpha_A)=0$
and $\beta_B(\alpha_B)=0$.
\item
In the case of theories approaching fixed-point behavior
$\beta_A(\bar\alpha_A)=0$ and
$\beta_B(\bar\alpha_B)=0$, the commensurate scale relation relates both
the ratio of
fixed point couplings $\bar\alpha_A/\bar\alpha_B$, and the ratio of
scales as the fixed point is approached.
\item
Commensurate scale relations satisfy the Abelian correspondence principle
\cite{BrodskyHuet};
\ie\ the non-Abelian gauge theory prediction reduces to Abelian theory for
$N_C \to 0$ at
fixed $ C_F\alpha_s$ and fixed $N_F/C_F$.
\item
The perturbative expansion of a commensurate scale relation has the same
form as a
conformal theory, and thus has no
$n!$  renormalon growth arising from the  $\beta$-function.
It is an interesting conjecture whether the perturbative expansion relating
observables to observable are in fact free of all $n!$ growth.  The
generalized Crewther relation, where the commensurate relation's perturbative
expansion forms a geometric series to all orders, has convergent behavior.
\end{enumerate}

Virtually any perturbative QCD prediction can be written in the form of a
commensurate
scale relation, thus eliminating any uncertainty due to renormalization
scheme or scale
dependence.  Recently it has been shown \cite{BPT} how the commensurate scale
relation between the radiative corrections to $\tau$-lepton decay and
$R_{e^+e^-}(s)$
can be generalized and empirically tested for arbitrary $\tau$ mass and nearly
arbitrarily functional dependence of the $\tau$ weak decay matrix element.

An essential feature of the \av(Q) scheme is the absence of any
renormalization scale ambiguity, since  $Q^2$ is,  by definition, the square of
the physical momentum transfer. The \av\ scheme naturally takes into
account quark mass
thresholds,  which is of particular phenomenological importance to QCD
applications in the mass region close to threshold.
As we have seen, commensurate scale relations provide
an analytic extension of the conventional \ms\  scheme in which many of
the advantages of the \av\ scheme are inherited by the \amst\ scheme,
but only minimal changes have to be made.
Given the commensurate scale relation connecting \amst\ to \av\, expansions in
\amst\ are effectively expansions in \av\ to the given order in perturbation
theory at a corresponding commensurate scale. Taking finite quark mass
effects into account analytically
in the running, rather than using a fixed flavor number $N_F$ between
thresholds, leads
to effects of the order of $1\%$ for the one-loop running coupling, with the
largest differences occurring near thresholds. These differences are important
for observables which are calculated neglecting quark masses, and could
turn out to be
significant when comparing  low and high energy measurements of the strong
coupling.

Unlike the conventional \ams\ scheme, the modified \amst\ scheme is
analytic at quark mass thresholds, and  it thus provides a natural
expansion parameter for perturbative representations of observables.
In addition, the extension of the \ms\ scheme, including quark mass
effects analytically, reproduces the standard treatment of  quark masses in the
\ms\ scheme to within a fraction of a percent. The standard treatment amounts
to either calculating multi-loop diagrams  with explicit quark masses or adding
higher dimension operators to the effective Lagrangian. These corrections
can be viewed as compensating for the fact that the number of flavors in the
running is kept constant between mass thresholds. By utilizing the BLM scale
setting procedure, based on the massless $N_F$ contribution, the analytic
extension  of the \ms\ scheme correctly absorbs both massless and mass dependent
quark contributions from QCD diagrams, such as the double bubble diagram, into
the running of the coupling. This gives the opportunity to convert
any calculation made in the \ms\ scheme with massless quarks into an
expression which includes quark mass corrections from QCD diagrams
by using the BLM scale and replacing \ams\ with \amst.

Finally, we note the potential importance of utilizing the  \av\
effective charge or the equivalent analytic \amst\ scheme in
supersymmetric and grand unified theories, particularly since the
unification of couplings and masses would be expected to occur in terms
of physical quantities rather than parameters defined by theoretical
convention.

\section*{Acknowledgments}
Much of this work is based on collaborations with Michael Melles, Mandeep
Gill, Hung Jung Lu, Andrei Kataev, and Gregory Gabadadze.
We also thank the organizers of RADCOR98, particularly Professor Joan
Sola of the Universitat Autonoma de Barcelona, for their outstanding arrangements
and hospitality.

\end{document}